\documentclass[%
reprint,
superscriptaddress,
 amsmath,amssymb,
 aps,
]{revtex4-2}

\usepackage{graphicx}
\usepackage{dcolumn}
\usepackage{bm}
\usepackage{xfrac}
\usepackage{comment}
\usepackage{mathtools}
\usepackage{physics}
\usepackage{makecell}
\usepackage{etoolbox}
\usepackage{xcolor}

\newcommand{\Ham}{\hat{H}}
\newcommand{\D}{\hat{D}}
\newcommand{\N}{\hat{N}}
\newcommand{\M}{\hat{M}}

\newcommand{\Id}{\hat{I}}
\newcommand{\dm}{\hat{\rho}}

\newcommand{\Lin}{\mathcal{L}}

\makeatletter
\patchcmd{\frontmatter@abstract@produce}
  {\vskip200\p@\@plus1fil
   \penalty-200\relax
   \vskip-200\p@\@plus-1fil}
  {}
  {}
  {}
\makeatother

\begin{document}

\preprint{}

\title{An experimental pathway towards an exact theory of strong coupling
}

\author{Eugenia Pyurbeeva}
\email{eugenia.pyurbeeva@mail.huji.ac.il}
\affiliation{The Institute of Chemistry and the Fritz Haber Center for Theoretical Chemistry, The Hebrew University of Jerusalem, Jerusalem 9190401, Israel}
\author{Ronnie Kosloff}
\affiliation{The Institute of Chemistry and the Fritz Haber Center for Theoretical Chemistry, The Hebrew University of Jerusalem, Jerusalem 9190401, Israel}

\date{\today}
\begin{abstract}
We employ a mathematically equivalent form of the GKLS master equation to arrive at an exact theoretical description of a two-level system strongly coupled to the environment. The framework, while intuitive, shedding light on the physics of the problem, and agreeing with existing results, such as thermalisation to a non-canonical state, is based around three parameters that are unknown outside of the weak coupling regime -- the analogue to the detailed balance relation, and two coupling strength constants. As a way forward, we propose a feasible experimental protocol based on a solid-state electronic quantum dot device, through which the fundamental parameters of the problem can be revealed, which would further the fundamental understanding of strong coupling. 
\end{abstract}
\maketitle
\begin{figure}
\includegraphics[width=\linewidth]{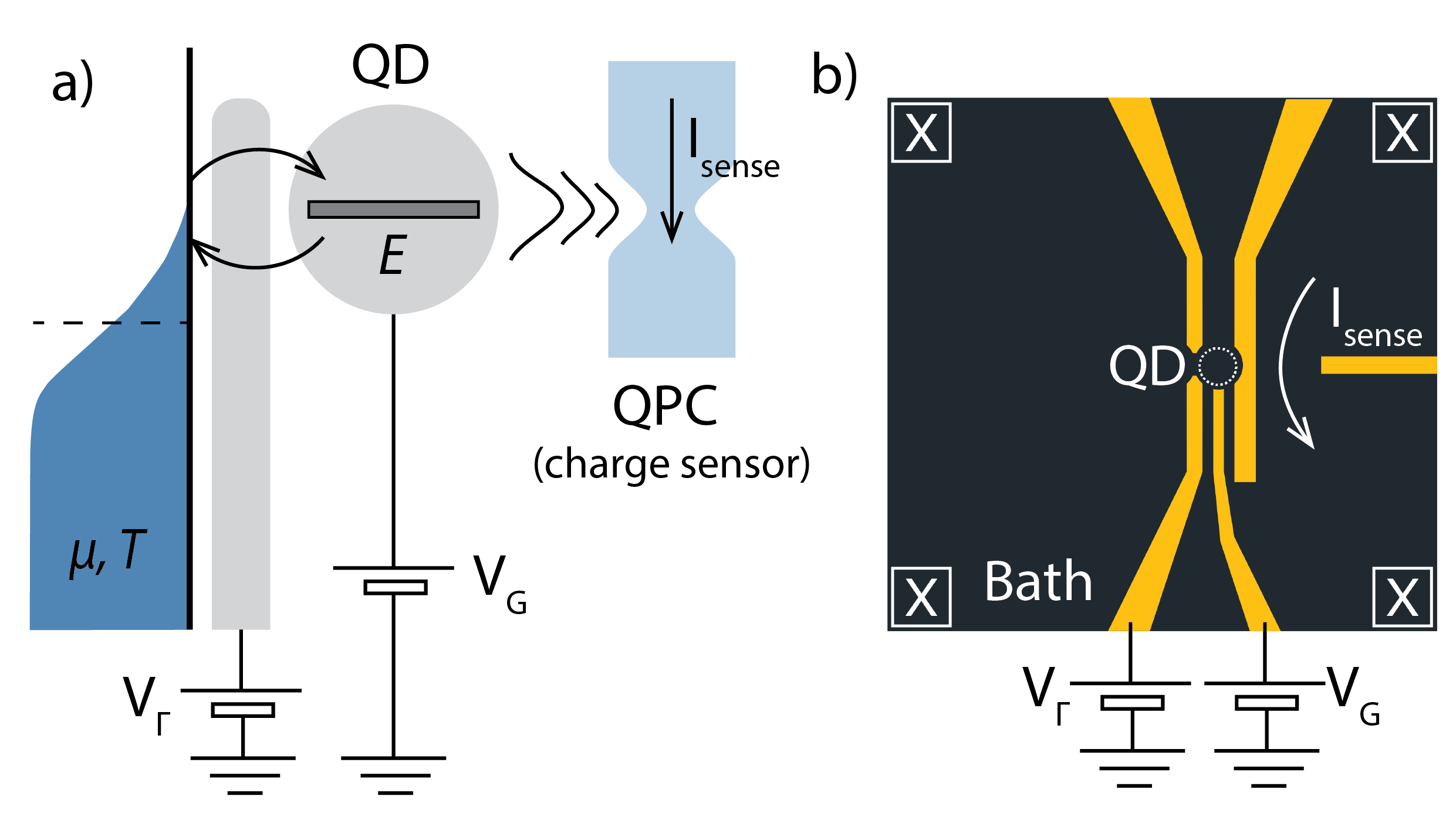}
\caption{a) The system setup. A quantum dot with addition energy $E$, controlled by a gate voltage $V_G$ exchanges electrons with a thermal bath, characterised by a temperature and chemical potential ($T$ and $\mu$). The coupling strength $\Gamma$ is controlled by a side-gate voltage $V_{\Gamma}$. A quantum point contact (QPC) located close to it acts as a charge sensor. b) A CAD layout of the gates for a device shown in panel (a).  
\label{fig:setup}}
\end{figure}

The theory of strong coupling is one of the Holy Grails in the field of open quantum systems. Strong interaction between the system and the bath can give rise to rich physics -- the build-up of correlations between the two \cite{Iles-Smith2014, Rivas2020}, non-Markovian effects \cite{Tiwari2025}, and thermalisation to a state deviating from a canonical distribution \cite{Iles-Smith2014, Burke2024}, to name a few. However, the strong coupling regime remains poorly understood, with many open questions, such as whether or not it can enhance the performance of thermal machines \cite{Gelbwaser-Klimovsky2015, Seah2018}. 

The reason for it lies in the nature of the tools of open quantum systems -- quantum master equations (QMEs), which give a reduced description of a subsystem coupled to a larger environment. Following an approach initiated by
Bloch, Redfield, and Fano \cite{wangsness1953dynamical, redfield1957theory, fano1957description}, most QMEs were derived assuming weak coupling between the system and its environment, and therefore fail to capture the strong coupling regime.

Extending the theory to include strong coupling is a subject of intense research with a multitude of approaches \cite{Anto-Sztrikacs2023}. Most, however, fall into one of two categories: phenomenological, describing the strong coupling as a Lorentzian broadening of the energy levels of the system; and modelling, where the size of the ``system'' is increased to include a part of the reservoir it is strongly coupled to, while the coupling to the rest of the environment is considered weak \cite{wang2015nonequilibrium,katz2016quantum}. The former, through its nature, does not allow for insight into physics, while the latter quickly becomes computationally intractable as the size of the enlarged system increases. 

We propose an alternative approach to the problem, based on experiment. Strong coupling effects, such as lifetime broadening, are routinely observed in nanoscale systems \cite{Thijssen2008}, however for the experiment to shed light on the underlying physics of the problem, one needs to construct a theoretical description based around a small number of experimentally accessible parameters that give insight into the system behaviour. This is the primary aim of the work.

To this end, as a paradigmatic example, we take a two-level system, a quantum dot with two energetically accessible charge states, coupled to an electron bath and exchanging electrons with it (Fig.\ref{fig:setup}). We derive a new, exact description of strong coupling in this case, based on three parameters, which, beyond the weak coupling limit, are generally unknown -- the modified detailed balance relation, and two coupling strength constants. We then demonstrate that they can be extracted from a realistic experiment in a quantum dot device shown in Fig.\ref{fig:setup}. 


The weak coupling approach \cite{davies1974markovian} taken in the initial derivations of QMEs is not unique. The Gorini--Kossakowski--Lindblad--Sudarshan (GKLS) master equation \cite{lindblad1976generators, gorini1976completely}:
\begin{equation}
  \label{eq:lindblad}
    \frac{\dd \dm}{\dd t}=-i [ \Ham, \dm]+\sum_i \gamma_i \Big(L_i \dm L_i^\dagger-\frac{1}{2}\{L_i^\dagger L_i, \dm\} \Big)  
\end{equation}
where $\dm$ is the density operator of the subsystem, while $\Ham$ and $L_i$ are arbitrary operators acting on the subsystem Hilbert space can be derived from a purely mathematical perspective, by demanding a completely positive and trace-preserving (CPTP) dynamical map \cite{kraus1971general}, which possesses at least one stationary (fixed-point) state \cite{frigerio1977quantum}. For Markovian dynamics, the rates $\gamma_i$ are non-negative, but the GKLS form can be generalized to describe non-Markovian dynamics by allowing the decay rates $\gamma_i(t)$ to become temporarily negative \cite{piilo2008non, dann2022non}. GKLS form (Eq.\ref{eq:lindblad}) can thus be thought of as a mathematical description of a much wider range of dynamics than those as weak coupling. 

In a recent work \cite{Pyurbeeva2026a}, we have shown that for a two-level system, Eq. \ref{eq:lindblad} can be written as a sum of free evolution and a combination of exchanges of ``generalised charges'', given by Hermitian operators $\N_i$, between the system and the bath(s), and dephasing terms, corresponding to external noise or weak measurement \cite{gorini1976n}. 

For a quantum dot with a single exchange process, the quantum master equation is:
\begin{equation}
\label{eq:exchange}
    \frac{\dd \dm}{\dd t}=-i [ \Ham, \dm]+\gamma_p \Lin_p(\dm, \N)+\gamma_m \Lin_m(\dm, \N)
\end{equation}
where the indices $p/m$ (plus/minus) in the rate coefficients and dissipator terms stand for the transfer of charge $\N$ to or from the system. 

Under strict energy conservation conditions, in the weak coupling regime, $\N$ in Eq.\ref{eq:exchange} has to commute with the system Hamiltonian, and the rate coefficients $\gamma_p$ and $\gamma_m$ obey detailed balance: $\gamma_p/\gamma_m=e^{-E/T}$, where $E$ is the difference between the energy levels of the two-level system \cite{Pyurbeeva2026}. However, the result in  \cite{Pyurbeeva2026a} was derived from the mathematical form of the GKLS equation employing no additional assumptions, and thus has no imposed restrictions.  

Here, we come to the central conceptual idea of our work: in the general case of a single exchange process for a system with no time-dependence, beyond the weak-coupling limit, the freedom in Eq.\ref{eq:exchange} lies in the non-commutation between $\Ham$ and $\N$. Therefore, the main feature of strong coupling lies in the commutator between the generalised charge and the Hamiltonian of the system, $[\Ham, \N]$, along with a new relation between $\gamma_p$ and $\gamma_m$ to replace detailed balance. This brings strong coupling into the realm of non-Abelian effects \cite{Guryanova2016, YungerHalpern2016, Lostaglio2017}, and gives a direct physical implementation of the exchange of non-commuting charges \cite{garwola2024open}. 

This approach aligns with physical intuition. For a system exchanging energy with the environment, the non-commutation between the Hamiltonian and the operator of the exchanged energy would lead to an uncertainty, with manifests in a broadened resonance peak -- a signature of strong coupling. Additionally, for a particle in two potential wells in the basis of $\ket{L}$, $\ket{R}$, $\sigma_z$ gives the well occupation, while $\sigma_x$ describes the coupling between the wells (which is real in the absence of magnetic field). Extending similar logic to a quantum dot with the basis states $\ket{0}$ and $\ket{1}$ for the QD occupation, $\sigma_z$ would correspond to the addition energy, while $\sigma_x$ and $\sigma_y$ would give the ``static'' and ``phase-dependent'' couplings.    

Following the results of \cite{Pyurbeeva2026, Pyurbeeva2026a} for the dissipator terms $\Lin_{p/m}(\dm, \N)$, Eq.\ref{eq:exchange} can be written as: 
\begin{multline}
\label{eq:master}
    \frac{\dd \dm}{\dd t}=-i E [ \Ham, \dm]-(\gamma_p+\gamma_m)(\dm-\frac{1}{2}\Id)+(\gamma_p-\gamma_m)\N+\\+(\gamma_p+\gamma_m)\frac{[\N, [\N, \dm]]}{2}
\end{multline}
where $\Ham$ (the normalised Hamiltonian of the system) and $\N$ are traceless $| \Ham |=| \N |=1/2$, so that the spacing between the eigenvalues is 1. The terms in this form represent, respectively: the free evolution; thermal mixing between the energy levels driving the system to equal populations; the drive towards alignment with $\N$; and thermal dephasing. Due to the linearity of the GKLS equation, external noise or weak measurement in $\hat{A}$ can be added as a double commutator term. 

We construct an orthonormal basis, such that
\begin{equation}
    \Ham=\cos{\varphi} \N +\sin{\varphi} \D
\end{equation} 
where $\D$ is orthogonal to $\N$, and $| \D |=1/2$ (this keeps the normalisation of $\Ham$). The third basis vector is $i [\N, \D]$. Then, $[\Ham, \N]=-i \sin{\varphi}  [\N, \D]$. The normalisation leads to double commutation relations:
\begin{equation}
    \begin{split}
    [\D, [\D, \N]]=\N\\
    [\N, [\N, \D ]]=\D       
    \end{split}
\end{equation}
Using these relations, we find the stationary state of Eq.\ref{eq:master} in the form:
\begin{equation}
    \rho=\frac{\Id}{2}+\beta \N + \alpha \D + i \lambda [\N, \D ] 
\end{equation}
arriving at:
\begin{equation}
    \begin{split}
    \beta=\frac{(\gamma_p-\gamma_m)}{(\gamma_p+\gamma_m)}\frac{4E^2\cos^2{\varphi}+(\gamma_p+\gamma_m)^2}{4E^2\cos^2{\varphi}+2E^2 \sin^2{\varphi}+(\gamma_p+\gamma_m)^2}\\
    \alpha=\frac{(\gamma_p-\gamma_m)}{(\gamma_p+\gamma_m)}\frac{4E^2\cos{\varphi}\sin{\varphi}}{4E^2\cos^2{\varphi}+2E^2 \sin^2{\varphi}+(\gamma_p+\gamma_m)^2}\\
    \lambda=(\gamma_p-\gamma_m)\frac{2E\sin{\varphi}}{4E^2\cos^2{\varphi}+2E^2 \sin^2{\varphi}+(\gamma_p+\gamma_m)^2}
    \end{split}
\end{equation}
To gain further physical understanding, we introduce two dimensionless parameters:
\begin{equation}
    \begin{split}
        \gamma=\frac{(\gamma_p-\gamma_m)}{(\gamma_p+\gamma_m)}
        \\
        \Gamma=\frac{(\gamma_p+\gamma_m)}{2E}
    \end{split}
\end{equation}
$\gamma$ relates to temperature, while $\Gamma$ has the meaning of effective coupling strength. Note that both $\Gamma$ and $\phi$ are measures of coupling, one from the rate coefficients and one from the non-commutation between $\Ham$ and $\N$, and therefore must be related. 

In the dimensionless form, the stationary state coefficients reduce to:
\begin{equation}
    \begin{split}
    \beta=\gamma \frac{\cos^2{\varphi}+\Gamma^2}{\cos^2{\varphi}+ \frac{\sin^2{\varphi}}{2}+\Gamma^2}\\
    \alpha=\gamma \frac{\cos{\varphi}\sin{\varphi}}{\cos^2{\varphi}+ \frac{\sin^2{\varphi}}{2}+\Gamma^2}\\
    \lambda=\gamma \frac{\Gamma\sin{\varphi}}{\cos^2{\varphi}+ \frac{\sin^2{\varphi}}{2}+\Gamma^2}
    \end{split}
\end{equation}
giving the stationary state as:
\begin{multline}
    \dm=\frac{\Id}{2}+\frac{\gamma}{\left(1- \frac{\sin^2{\varphi}}{2}\right)+\Gamma^2}\left((\cos^2{\varphi}+\Gamma^2)\N+\right.
     \\\left.+ \cos{\varphi}\sin{\varphi}\D +i \Gamma \sin{\varphi} [\N, \D] \right)
\end{multline}
Or, returning to $\Ham$, $\N$, and $[\Ham, \N]$,
\begin{equation}
\label{eq:rho-0}
    \dm=\frac{\Id}{2}+\frac{\gamma}{\left(1- \frac{\sin^2{\varphi}}{2}\right)+\Gamma^2}\left(\cos{\varphi} \Ham -i \Gamma [\Ham, \N] + \Gamma^2\N \right)
\end{equation}
While, for simplicity, it is not in the standard exponential form of a Gibbs state, the matrix exponent in two dimensions is linear by the operator. What is meant by it is that:
\begin{equation}
    \dm_\text{Gibbs}=\frac{e^{\M}}{\Tr \left(e^{\M} \right)}=\frac{\Id}{2}+ \left( \frac{1-e^{|\M|}}{1+e^{|\M|}}\right) \frac{\M}{|\M|}
\end{equation}
This means that, aside from the scaling, the non-identity part of the stationary state in Eq.\ref{eq:rho-0} can be analysed as the exponentiated term in the Gibbs distribution. It brings several interesting and noteworthy insights. First, as noticed in \cite{Pyurbeeva2026a}, in comparison to the usually written form of the Gibbs state for a grand-canonical ensemble \cite{Site2024, Reible2025}:
\begin{equation}
    \dm_{MC}=\frac{1}{Z}e^{-\beta (\Ham - \mu \N)}
\end{equation}
there is an additional term with the commutator $[\Ham, \N]$, which isn't typically considered, as $\N$ is assumed to commute with $\Ham$. Besides this, the coefficient with $\N$ in Eq.\ref{eq:rho-0} is proportional to $\Gamma^2$, and thus can not have the meaning of chemical potential. Instead, $E-\mu$ has to come out in the coefficient $\Ham$, the normalised Hamiltonian, which means that the chemical potential acts on the projection of $\N$ on $\Ham$. This is physically reasonable, as $\mu$ is an energetic quantity, and agrees with the widely known notion that $E-\mu$ acts as an effective energy in particle exchange \cite{Pyurbeeva2023, Reible2025}. Finally, as a function of $\Gamma$, when $\Gamma\rightarrow 0$, the stationary state (Eq.\ref{eq:rho-0}) is aligned with $\Ham$, while at strong coupling, when $\Gamma^2\gg \Gamma \gg 1$, particle exchange dominates over free evolution and the stationary state aligns with $\N$. In the intermediate regime, the commutator $[\Ham, \N]$ plays a significant role. 

The considerations above support the tenability of the non-Abelian representation of strong coupling. However, as the framework is based on purely algebraic manipulations of the GKLS equation, it does not introduce new physics. The main physical question following the result in Eq.\ref{eq:rho-0} are the relations between the effective detailed balance $\gamma$ and the two coupling constants $\Gamma$ and $\phi$. 

While we can not offer a theoretical approach to the problem, we suggest an experimental method for studying these relations. The experimental setup is based on a single gate-defined quantum dot in the Coulomb blockade regime (Fig.\ref{fig:setup}a) exchanging electrons with a Fermi bath. The coupling to the bath and the addition energy can be controlled by a plunger gate (with $V_G$) and an entrance gate (with $V_\Gamma$) respectively. A quantum point contact (QPC) \cite{VanWees1988} positioned next to the quantum dot acts as a charge sensor. Devices of this kind (Fig.\ref{fig:setup}b) have been routinely studied since the late 80s \cite{VanWees1988, Kastner1992, Lindemann2002}, and more recently, this exact configuration has been used in direct entropy measurement experiments \cite{Hartman2018, Pyurbeeva2022, Kealhofer2025}. 

Besides the control of addition energy and coupling strength, and strong measurement of the quantum dot occupation, the experimental setup allows for the addition of two sources of dephasing: one through weak measurement of $\N$ (by passing a small current through the QPC); and another through applying white noise in voltage to either the bath (fluctuating $\mu$), or, equivalently, all the gates defining the quantum dot (fluctuating the entire potential well profile $\Ham$). These add, respectively, the dephasing terms $\gamma_N [\N, [\N, \dm]]/2$ and $\gamma_H [\Ham, [\Ham, \dm]]/2$ to the master equation (Eq.\ref{eq:master}).

As the only component of the density matrix that can be measured is $\N$, we find $\beta_N$ and $\beta_H$ -- the coefficients with $\N$ for the stationary state under the application of external noise in $\N$ and $\Ham$ as a function of the noise strength. 

For the weak measurement in $\N$:
\begin{equation}
    \beta_N=\gamma \frac{\cos^2{\varphi}+(\Gamma+\Gamma_N)^2}{\cos^2{\varphi}+\frac{\sin^2{\varphi}}{2}\left(1+\frac{\Gamma_N}{\Gamma} \right)+(\Gamma+\Gamma_N)^2}
\end{equation}
while for the fluctuations of $\Ham$:
\begin{equation}
    \beta_H=\gamma \frac{\cos^2{\varphi}-\Gamma_H(\Gamma+\Gamma_H)+(\Gamma+\Gamma_H)^2}{\cos^2{\varphi}+\frac{\sin^2{\varphi}}{2}-\frac{\Gamma_H(\Gamma+\Gamma_H)}{2}\sin^2{\varphi}+(\Gamma+\Gamma_N)^2}
\end{equation}
where we have defined $\Gamma_N=\frac{\gamma_N}{2E}$, $\Gamma_H=\frac{\gamma_H}{2E}$. 

The three central parameters of the problem, $\gamma$, $\Gamma$, and $\varphi$ can be found as follows:
\begin{enumerate}
    \item In the limit of continuous strong measurement of $\N$ (it has to be noted that continuous strong measurement affecting system dynamics differs from a one-off strong measurement leading to the collapse of the quantum state):
    \begin{equation}
        \lim_{\Gamma_N \to \infty} \beta_N = \gamma 
    \end{equation}
    This allows to find $\gamma$. It also agrees with the result for the mean population found from the classical rate equation, disregarding any quantum effects. 
    \item Under strong potential noise:
    \begin{equation}
        \lim_{\Gamma_H \to \infty} \beta_H = \gamma \frac{\cos^2{\varphi}}{1-\frac{\sin^2{\varphi}}{2}}
    \end{equation}
    Comparing with the previous result allows to find $\varphi$. 
    \item Finally, in the absence of external noise, the measurement of $\N$ (as per Eq.\ref{eq:rho-0}):
    \begin{equation}
        \beta_0=\gamma \frac{\cos^2{\varphi}+\Gamma^2}{\cos^2{\varphi}+ \frac{\sin^2{\varphi}}{2}+\Gamma^2}
    \end{equation}
    allows to determine $\Gamma$, the final of the three relevant parameters. 
\end{enumerate}
\begin{figure}
\includegraphics[width=\linewidth]{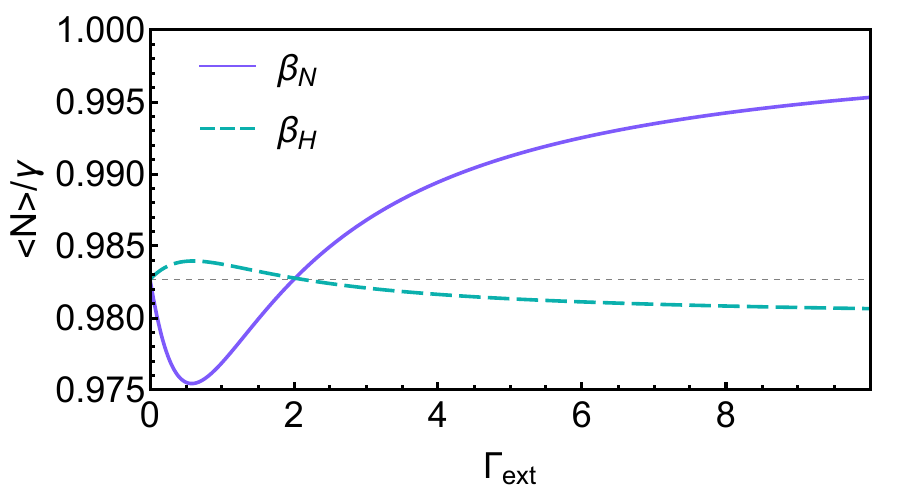}
\caption{A sample dependence of mean population in the presence of external dephasing in $\N$ and $\Ham$ as a function of noise intensity, $\Gamma_\text{ext}$, standing for $\Gamma_N$ or $\Gamma_H$. In the case plotted, $\Gamma=0.4$, $\varphi=0.2$. 
\vspace{-0.5cm}
\label{fig:plot}}
\end{figure}
Figure \ref{fig:plot} shows the dependence of the mean population of the quantum dot, $\beta_N$ and $\beta_H$ as a function of the external dephasing strength $\Gamma_N$ or $\Gamma_H$, shown as $\Gamma_\text{ext}$. In the weak noise regime the dephasing in $\N$, as expected, leads to a lower mean population, while the dephasing in $\Ham$ increases it. The latter is due to the fact that dephasing destroys the components of the density matrix orthogonal to the dephasing operator, which, in the case of $\Ham$ are $i[\Ham, \N]$ and $-\sin{\varphi} \N + \cos{\varphi} \D $. The negative sign in $\N$ in the second vector leads to a restoration of mean population. In the strong dephasing regime, however, for the noise in $\N$ the destruction of the density matrix components orthogonal to $\N$ leads to an increase of mean $N$, while for the noise in $\Ham$ the mean population is overall lower than in the absence of external noise. While the complete measurement of $\beta_N(\Gamma_N)$ and $\beta_H(\Gamma_H)$ is not necessary to determine the parameters of the theory, and the limits are sufficient, the full measurement can act as further test of the validity of the $\Ham$ and $\N$ non-commutation model. 

The main complication of the experimental realisation of the suggested protocol lies in the implementation of strong projecting measurement. We have assumed it to be instantaneous, however for such a measurement to be performed, sufficient the QPC needs to have a sufficient voltage across it for the two charge states of the quantum dot to be distinguishable. Ramping this operating voltage up introduces a new timescale into the system, and it needs to be much shorter than the characteristic lifetime of the electron on the quantum dot, which limits the ability to conduct quantum measurements at strong coupling. This detail, however, is purely practical, and we leave it for future experimental work. Finally, we note than the proposed experiment is not restricted to electronic systems, and a similar measurement can be performed on a single qubit. 

To conclude, we suggest a novel simple theoretical approach to the problem of strong coupling, based on the non-commutation between the system Hamiltonian and the quantity exchanged between the system and the bath, which casts strong coupling as a non-Abelian effect. The theoretical framework agrees with physical intuition, giving the expected effect of a finite resonance width, and also suggest new effects, such as the presence of the $[\Ham, \N]$ commutator term in the generalised Gibbs distribution, and the number operator $\N$ in it being modified by the square of the coupling strength, as opposed to the chemical potential. 

The theory is based on purely mathematical manipulation of the GKLS equation, and therefore is general, however, as a consequence, it lack the predictive ability for new physics. Instead, we suggest an experimental protocol, based on charge sensing in a gate-defined quantum dot with controlled coupling strength to the bath, which allows to determine the parameters of the model. This will allow the experiment to shed new light on the theory of strong coupling -- the modification of detailed balance and the relation between coupling strength and the hybridisation of the electrons between the bath and the system.

Finally, beyond the immediate results presented in this work, we hope it opens a new, experimental, direction for study of open quantum systems. We have shown that even a well-established device has the potential to yield profound insight into fundamental questions in the field. The current advances in nanotechnology offer a wealth of quantum systems in which answers can be revealed.

\section*{Acknowledgements}
The authors thank Declan Mahony and Mark Blumenthal for the image of a device for suitable for the experimental realisation of the work. \\ 
E.P. is grateful to the Azrieli Foundation for the award of an Azrieli Fellowship.





\providecommand{\noopsort}[1]{}\providecommand{\singleletter}[1]{#1}%

\end{document}